\documentclass[10pt]{iopart}
\usepackage[utf8]{inputenc} 
\usepackage{graphicx}
  \expandafter\let\csname equation*\endcsname\relax
  \expandafter\let\csname endequation*\endcsname\relax
\usepackage{amsmath}
\usepackage{amsfonts}
\newcommand{\be}{\begin{equation}}
\newcommand{\ee}{\end{equation}}
\newcommand{\bea}{\begin{eqnarray}}
\newcommand{\eea}{\end{eqnarray}}

\usepackage[margin=1.25in]{geometry}

\newcommand{\ket}[1]{\left|#1\right\rangle}

\usepackage{fancyhdr}
\usepackage{color}

\begin{document}
\title{An adjustable-length cavity and Bose-Einstein condensate apparatus for multimode cavity QED}
\author{Alicia J. Koll\'{a}r\footnote{These authors contributed equally.}, Alexander T. Papageorge\footnote{These authors contributed equally.}, Kristian Baumann, Michael A. Armen, and Benjamin L. Lev}
\address{Department of Applied Physics, Stanford University, Stanford CA 94305, USA}
\address{Department of Physics, Stanford University, Stanford CA 94305, USA}
\address{E. L. Ginzton Laboratory, Stanford University, Stanford CA 94305, USA}
\ead{benlev@stanford.edu}

\date{\today}

\begin{abstract}
We present a novel cavity QED system in which a Bose-Einstein condensate (BEC) is trapped within a high-finesse optical cavity whose length may be adjusted to access both single-mode and multimode configurations.  We demonstrate the coupling of an atomic ensemble to the cavity in both configurations.  The atoms are confined either within an intracavity far-off-resonance optical dipole trap (FORT) or a crossed optical dipole trap via transversely oriented lasers.  Multimode cavity QED provides fully emergent and dynamical optical lattices for intracavity BECs.  Such systems will enable explorations of quantum soft matter, including superfluid smectics, superfluid glasses, and spin glasses as well as neuromorphic associative memory.
\end{abstract}

\tableofcontents

\pagestyle{fancyplain}
\lhead{}
\rhead{}

\section{Introduction}\label{intro}

Cavity QED is a particularly attractive setting  to explore many-body physics using quantum gases of neutral atoms~\cite{Ritsch13}.  In these systems, a laser pumps energy into cavity modes either directly via a semitransparent mirror or indirectly via coherent (Rayleigh) scattering off intracavity atoms. This light leaks out of the cavity by the  same semitransparent mirrors---or via atomic spontaneous emission---and may serve as a record of the collective dynamics driven by light-matter coupling.   Interaction timescales can be orders of magnitude faster than the superexchange or dipolar exchange interactions exploited in traditional optical lattice-based quantum simulators~\cite{Bloch:2008td,Cirac12,Yan:2013fn}. A non-equilibrium steady state may be reached before spontaneous emission heats the gas.   Despite the driven, dissipative character of this intrinsically non-equilibrium system, statistical mechanical treatments may be employed to analyze phase transitions~\cite{ritschpra,Littlewood06,Gopalakrishnan09,Gopalakrishnan10,Torre13}.  
  
In the dispersive limit, collective dynamics are driven by the effective atom-atom coupling mediated by the exchange of photonic excitations via the resonant cavity mode(s).   These atom-atom interactions are of infinite range in a single-mode Fabry-Per\'{o}t cavity~\cite{Rempe00}:  Mean-field descriptions are adequate to describe the collective behavior since spatially inhomogeneous interaction fluctuations are not supported, aside from an underlying sinusoidal modulation. By contrast, multimode Fabry-Per\'{o}t cavities---i.e., those with many degenerate modes---support shorter-ranged and strongly fluctuating interactions, as described in section~\ref{prospects}, and may enable observations of beyond mean-field many-body physics.  Quantum liquid crystals and superfluid glasses~\cite{Gopalakrishnan09,Gopalakrishnan10}, as well as spin glasses~\cite{Gopalakrishnan:2011jx,Strack:2011hv,Muller12,Buchhold13}, may be created by scattering light into a multimode cavity via an intracavity quantum gas. Spin glasses are intimately related to architectures for neuromorphic computation, specifically Hopfield associative memories~\cite{Gopalakrishnan:2011jx,Gopalakrishnan:2011cf} and 3D Ising machines~\cite{Marandi:2014ik}, both of which may be created in this quantum optical setting. 

Benchmarking observations of multimode cavity QED against prior, single-mode cavity experiments will prove crucial for understanding how complex beyond mean-field physics is manifest, if at all.  We have therefore constructed a cavity QED apparatus that provides a unique capability:  \textit{in situ} tunability of the cavity length from the single-mode regime to multimode degeneracy, all while maintaining stability against vibration in a vacuum sufficient for quantum gas production.

We first describe the experimental apparatus, beginning with 
the adjustable-length and multimode cavity in section.~\ref{cavity}, including vibration support, length adjustment, locking scheme, and cavity properties.   Section~\ref{BEC} describes the  portions of the apparatus used to produce ultracold gases, transport them into the cavity, and Bose-condense them or trap them using an intracavity far-off-resonance optical dipole trap (FORT)~\cite{Grimm00}.  Section~\ref{coupling} presents data on atomic ensemble-cavity coupling in both the single-mode and multimode configurations.    Lastly, we  briefly outline in section~\ref{prospects} how this adjustable-length and multimode cavity system might enable the  creation of exotic many-body states.

\section{Adjustable-length and multimode cavity}\label{cavity}

The challenge to engineering an adjustable-length optical cavity compatible with Bose-Einstein condensates (BECs) and capable of accessing the aforementioned regimes of many-body physics is many-fold, for instance: 1) All parts---including epoxies---of the cavity assembly and vibration isolation system must be compatible with maintaining a ultrahigh vacuum (UHV), or else BEC production will not be possible.   2) Magnetic parts must be avoided to achieve precise control of the magnetic fields and gradients at the position of the intracavity atoms and to avoid inducing vibrations during rapid switching of current in nearby electromagnetic coils. 3) The long-throw actuator that moves one of the mirrors must be sufficiently stable so as not to subject the cavity to too many low-frequency mechanical resonances.  Any such resonances allow ambient acoustical or seismic noise to drive the cavity below the low-pass cut-off of the filter provided by the  mechanical vibration isolation crossed.  Canceling these transmitted vibrations via feedback to a fast, single-crystal piezo, e.g., attached to the opposing mirror, is difficult:  large gain is required to compensate for the resonantly coupled noise, but that also makes it challenging to close the servo bandwidth before actuator resonances cause positive feedback oscillations.   4) To achieve a high-density of degenerate modes at the length of multimode operation, the mirror mounting procedure must provide the flexibility to angle and translate the mirrors as they become permanently affixed as the epoxy sets.  5) To maintain high-quality resonances throughout the travel of the mirror, the actuator must move with high linearity so as not to misalign the mirror.  6) The mirror spacing must be as small as possible while maintaining a large cavity finesse.   This provides a  large vacuum (single-photon) Rabi  frequency $2g_{0}$, which in turn maximizes the per-atom cooperativity, $C\equiv 2g_{0}^{2}/{\kappa\gamma}$~\cite{Kimble1998,TanjiSuzuki:2011dk}.  This parameter sets the strength of the atom-cavity interaction versus decay rates, where $\gamma$ is the atomic linewidth and the field decay rate is $\kappa$.
We now describe how we meet these challenges.

\begin{figure}[b]
\centering
\includegraphics[width=0.6 \columnwidth]{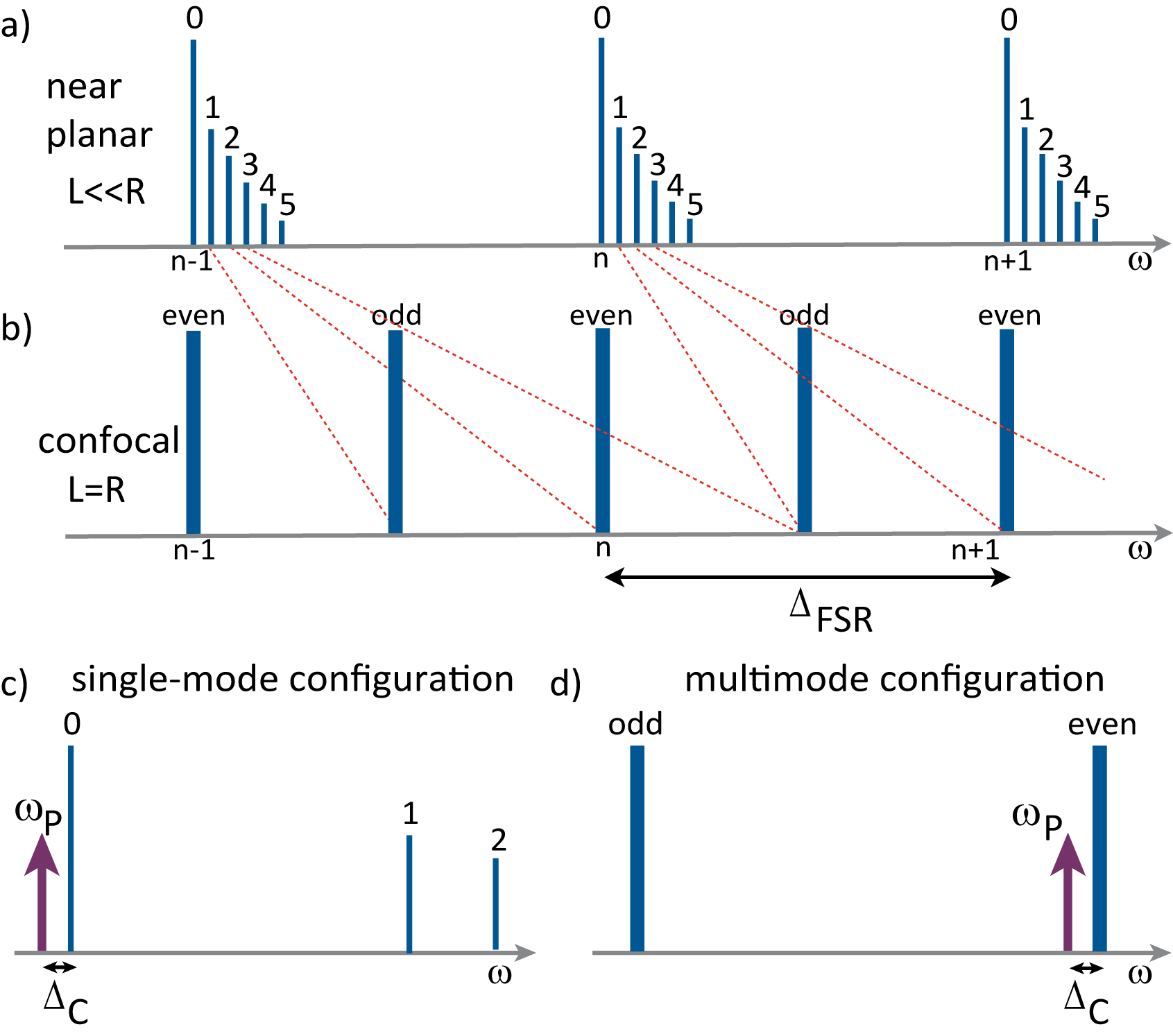}
\caption{Schematic cavity transmission spectra for single and multimode cavities. a) Transmission spectrum for a near planar cavity. TEM$_{0,0}$ modes are separated by $\Delta_\text{FSR}$. Higher order transverse modes are slightly shifted in frequency from the TEM$_{0,0}$ and tend to have weaker spectral intensity due to poor mode matching to pump. b) Transmission spectrum for a multimode, confocal cavity. Higher order transverse modes are shifted with respect to the TEM$_{0,0}$ by  $\Delta_\text{FSR}/2$  per transverse index $m$ or $l$. As a result, each TEM$_{0,0}$ mode is degenerate with all higher-order modes TEM$_\text{l,m}$ that possess a sum of transverse indices $l+m$ that is even. The odd higher-order modes are also degenerate but $\Delta_\text{FSR}/2$ away in frequency. c) Schematic of off-resonant transverse pumping in the single-mode case. The pump laser is located in close proximity to one particular cavity resonance. All other modes are insignificantly coupled due to their much larger detuning. d) Schematic of off-resonant transverse pumping in the case of the confocal cavity. The pump laser is detuned from the cavity resonance, but it is equally detuned from all TEM$_{0,0}$ and  higher-order even modes comprising the nearby family.} \label{cavityspect}
\end{figure}

\subsection{Multimode versus single-mode cavities}

 A single-mode Fabry-Per\'{o}t cavity possesses a geometrical configuration of mirrors such that families of transverse electromagnetic modes TEM$_{l,m}$   differing in the sum $l+m$ resonate at intervals in frequency much larger than $\kappa$.  For example, the frequency spacing between the TEM$_{0,0}$ mode and the family of $l + m =1$ of modes, i.e., TEM$_{1,0}$ and TEM$_{0,1}$, must satisfy $|\Delta\omega_{\{l+m=0\}-\{l+m=1\}}| \equiv |\omega_{\{l+m=0\}} - \omega_{\{l+m=1\}}|  \gg \kappa$, where $\omega_{\{l+m\}}$ is the frequency of all the modes in a family whose transverse field nodes sum to $l+m$. (In the absence of astigmatism and birefringence, $\omega_{1,0}=\omega_{0,1}$.)   The cavity is said to be in a  single-mode configuration with respect to a pump field  if the frequency detuning $\omega_{P}-\omega_{C} = \Delta_{C}$ of the pump laser from the single cavity mode of interest is much smaller than the frequency spacing between all other modes, e.g., if $| \Delta_{C} |\ll \Delta\omega_{\{l+m=0\}-\{l+m=1\}} $.  See figure~\ref{cavityspect} for  sketch.

Unlike single-mode cavities, multimode cavities support   TEM$_{l,m}$ modes resonating at the same frequency, even if they possess differing numbers $l+m$ of transverse field nodes~\cite{siegman}.   Interactions mediated by these modes are energy degenerate, since they possess the same frequency.  However, the spatial range of the effective interaction  is reduced due to the exchange of photonic excitation though superpositions of these disparate mode patterns:  Interferences among the mutually incommensurate Hermite-Gaussian---or equivalently, Laguerre-Gaussian---mode-functions $\Xi_{l,m}$ reduce the interaction range.  Specifically, the infinite-range interactions in a single-mode cavity form a complete graph of pairwise interactions weighted by the product of a single-field mode amplitude $\Xi_{l,m}(r_{i})\Xi_{l',m'}(r_{j})\delta_{ll'}\delta_{mm'}$.  By contrast, the finite-range interactions in a multimode cavity fall with distance as $\sin(\zeta\rho_{ij})/\zeta\rho_{ij}$. This interaction approaches a $\delta$-function in the limit $\zeta\rightarrow\infty$.  Here, $\zeta$ is the number of degenerate cavity modes, $r_{i}$ is the position of the atom $i$, and $\rho_{ij}$ is the separation between two atoms in the direction transverse to the cavity axis.  While shorter-ranged in the transverse plane, the interaction range remains infinite and proportional to $\sin(k_{0}z)=\sin(2\pi z/\lambda)$ along the cavity axis for the atomic cloud length and cavity geometry considered here. $\lambda$ is the wavelength of the cavity resonance.  

Multimode Fabry-Per\'{o}t cavities possess special relationships between the cavity length $L$ and the mirror radius of curvature $R$ to ensure that the round-trip path-length of rays corresponding to distinct wavevectors $\mathbf{k}$ are equal.  Common configurations are the concentric cavity $L=2R$ and the confocal cavity $L=R$.  The mirror foci of concentric cavities meet at the cavity center; the mirrors of typical concentric cavities  trace out but a small solid angle of a sphere of diameter $2R$.  One can see that rays with $\mathbf{k}$'s of equal length but differing direction reflect onto themselves, and therefore modes with spot sizes smaller than the mirror diameter are simultaneously resonant. (In practice, this spot size is limited by aspheric aberration and poor mirror coating reflectivity or high absorption near the mirror edge.)  The advantages of concentric cavities include small mode waists, providing larger $C$ compared to other equal-length resonators configurations~\cite{Vuletic01}. While concentric cavity  modes are describable by simple Bessel functions, the condition $L=2R$ lies at the instability boundary of two-mirror resonators~\cite{siegman}, forcing one to set the mirror length just shy of concentriccity, $L = 2R-\epsilon$, to ensure stability, in practice.  

Confocal cavities with identical mirrors, by contrast, lie between stable resonator regimes, and so their length may be continuously tuned through the multimode configuration point $L=R$.  Either Hermite-Gaussian or, equivalently, Laguerre-Gaussian transverse mode functions may be used to form a complete basis of supported modes~\cite{siegman}.   In a ray-tracing picture, each degenerate mode traces an hour-glass, bow-tie pattern~\cite{ChanThesis03}.   The frequency interval---free-spectral range---between sets of modes of equal parity is $\Delta_\text{FSR}=c/2L$, with that between  sets of opposite parity  $\Delta_\text{FSR}/2$.  See figure~\ref{cavityspect}.  This unusual property will allow us to couple the atoms to purely even parity modes or purely odd parity modes.   The degenerate modes of confocal cavities, while more difficult to work within the framework of theories employing functional integration~\cite{Gopalakrishnan09,Gopalakrishnan10}, provide access to the same physics as concentric cavities.  We have therefore chosen to build a cavity of the confocal configuration, though our system may easily be modified---without breaking vacuum in the ultracold atom production chamber---to increase the cavity length to the concentric configuration. 

\subsection{Cavity science chamber}

\begin{figure}[t!]
\centering
\includegraphics[width=1 \columnwidth]{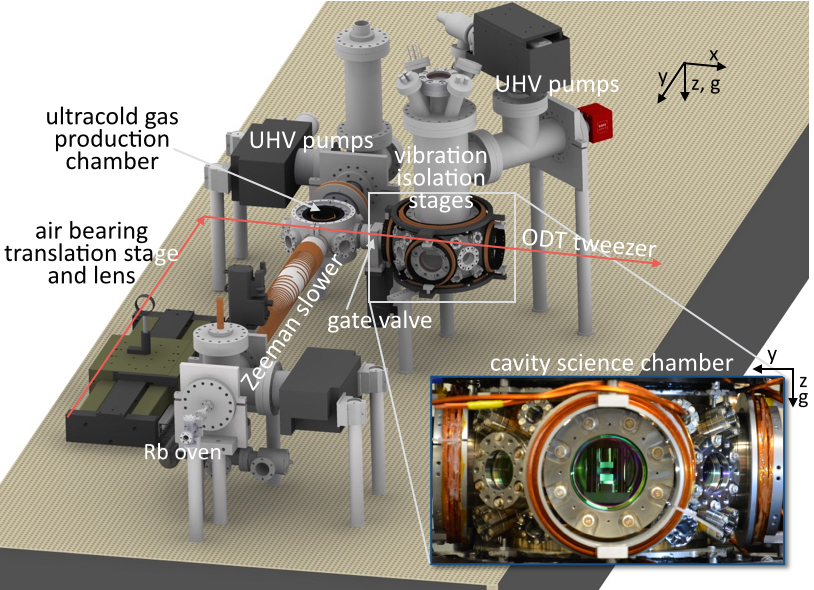}
\caption{Computer rendering of experimental apparatus, excluding laser optics. Inset: photograph of multimode cavity within science chamber.} \label{apparatus}
\end{figure}

\begin{figure}[t!]
\centering
\includegraphics[width=1 \columnwidth]{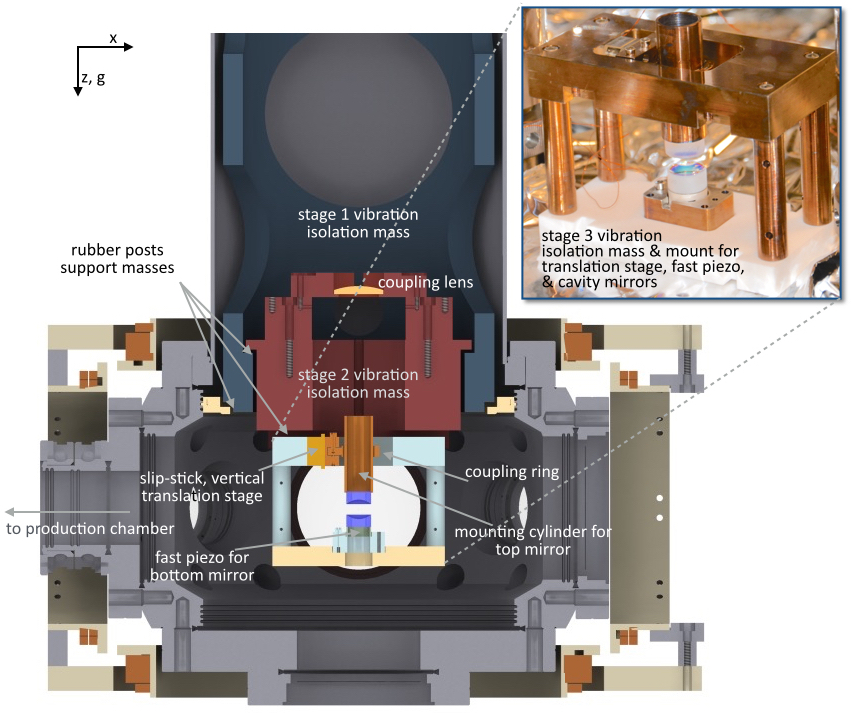}
\caption{Computer rendering of a section-view of the science chamber showing the cavity mount, vibration isolation stages, and in-vacuum coupling lens. Inset: photograph of cavity holder outside of vacuum, including single-crystal PZT, mirrors, and slip-stick translation stage.} \label{cavitychamber}
\end{figure}

Our experimental apparatus, shown in figure~\ref{apparatus}, consists of two sections: an ultracold atom production section and a cavity science chamber.  The science chamber, shown in detail in figure~\ref{cavitychamber}, houses the multimode cavity and the three stages of its vibration isolation stack.  This section provides ample optical access for coupling light into and out of the cavity, for trapping the atoms within the cavity using various configurations of optical dipole traps (ODTs), and for imaging the intracavity atomic cloud. 

\begin{figure}[t!]
\centering
\includegraphics[width=1 \columnwidth]{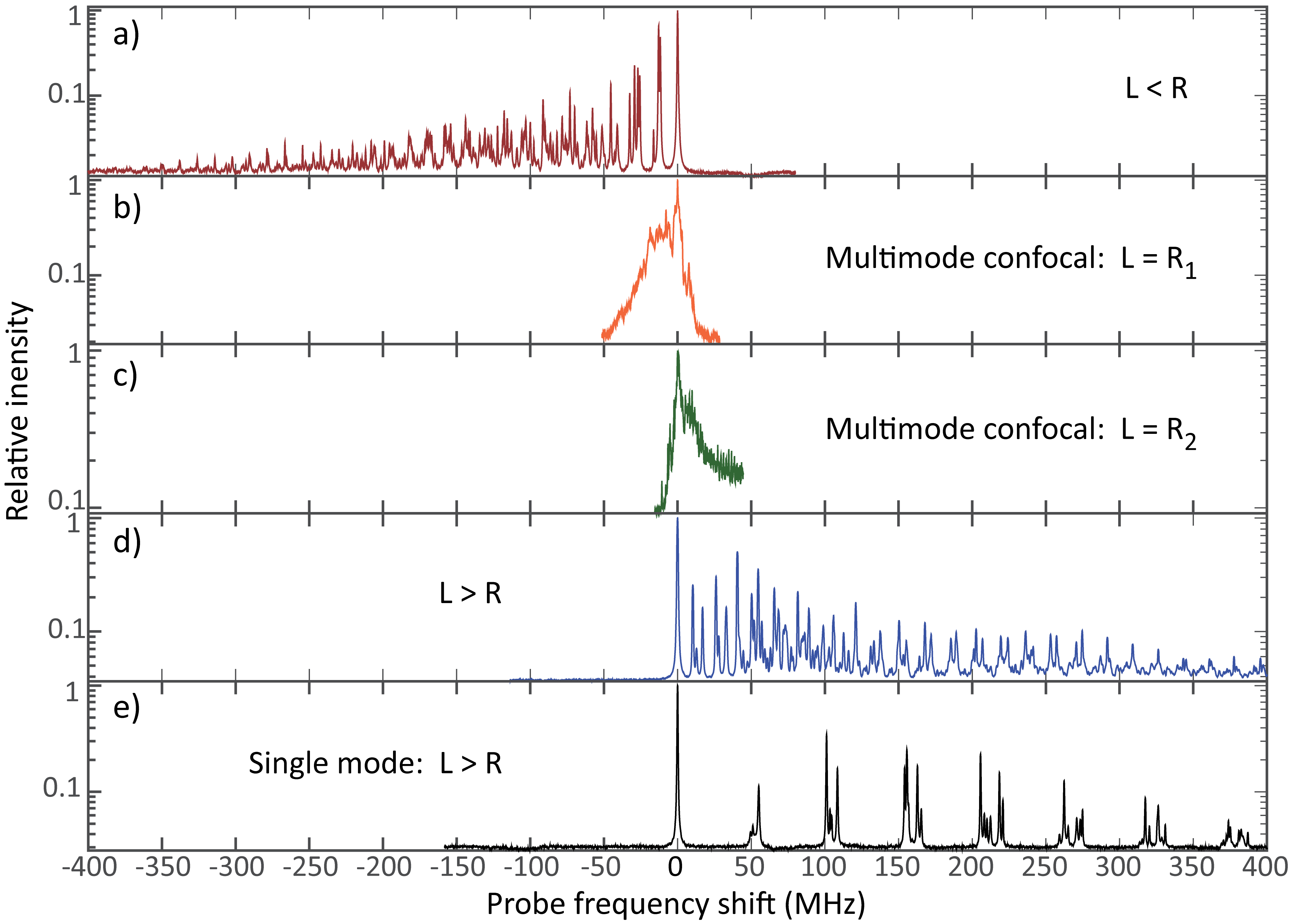}
\caption{
Cavity transmission spectra versus frequency for multiple cavity lengths.
a) Non-degenerate cavity with $L = R_1 - 9.6(1.0) \ \rm{\mu m}$, where the mirror radius of curvature along one of the two astigmatic axis is $R_1 = 9.959(1)$~mm and $R_2 =  R_1 + 8.6(1.1)$ $\mu$m.
Astigmatism in the cavity results in a splitting of  mode families with constant $l+m$. Since $L <R_1,R_2$, higher-order transverse modes appear at lower frequencies.
b) Confocal cavity with $L \approx R_1$. Even $l$ modes with $m=0$ are degenerate to within a few cavity linewidths, and modes with $m \neq 0$ form a broad shoulder extending out to lower  frequencies due to astigmatism.
c) Confocal cavity with $L \approx R_2$.  Even $m$ modes with $l = 0$ are degenerate to within a few cavity linewidths, and modes with $l \neq 0$ form a shoulder extending out to higher frequencies.
d) Non-degenerate cavity with $L = R_2 + 13.3(6) \ \rm{\mu m}$. 
Since $L >R_1,R_2$, higher-order transverse modes appear at higher frequencies.
e) Single-mode cavity regime with $L = R_2 + 53.4(6) \ \rm{\mu m}$, and lowest order families of constant $l+m$ modes are separated by approximately $50\ \rm{MHz}$.  Not shown, but even greater splittings between mode families is easily  achieved via additional extension of the cavity length.  We note that the resonance peak heights are highly dependant on the spatial structure of the in-coupled pump beam.} 
\label{modetrain}
\end{figure}

The vertically oriented mirrors of the high-finesse Fabry-Per\'{o}t cavity are affixed to a mount constructed from non-magnetic stainless steel (SS) and Macor.  This mount rests on a vibration-isolation crossed within the UHV chamber for the purpose of damping environmentally coupled vibrations. Length stabilization is predominantly effected by a single-crystal PZT on which the lower mirror is affixed with ultrahigh vacuum-safe epoxy. The upper mirror is similarly epoxied to one end of a stainless steel tube which is then epoxied to a SS mounting ring.  This ring is then bolted onto the slip-stick piezo positioning stage.  The stage is screwed to the stainless steel-and-Macor cavity mount and contains a slip-stick piezo actuator with 3~mm of travel along the cavity axis.  While non-repeatable, the tiny differential distance covered by each incremental actuation of the slip-stick piezo may easily be compensated by applying a small DC voltage to the stack piezo forming the very same actuator.  Continuously tuning the length of the cavity by hundreds of microns is straightforward and travel is sufficiently linear so as to be able to observe spectra of well-coupled modes throughout this travel range, see figure~\ref{modetrain}. 

The rectangular cavity mount provides ample optical access, allowing beams to be directed at the atoms from any of the twenty-four viewports mounted on the spherical octagon UHV chamber.  The mount hangs from the hooks of a heavy (3.5 kg) steel holder forming the second vibration isolation stage.  This rests on the internal rim of a heavy (6 kg) steel cylinder forming the first vibration isolation stage.  This cylinder then rests on a ledge around the inside of the UHV chamber. Small, several-mm-cubed, Viton rubber pieces separate the metal pieces at each interface, thereby forming a three-stage low-pass vibration filter. The entire vibration-isolation stack can be modeled as a six-pole low-pass filter for vibrations transferred from the chamber to the cavity. Simulations indicate that all resonances lie below 200~Hz and that the cavity mount is well-protected from environmental vibrations above this cutoff.  Figure~\ref{cavitychamber} shows a schematic of the primary vibration isolation stack pieces. Not shown are the ion  and non-evaporative getter  pumps connected to the cavity science chamber, though these may be seen in figure~\ref{apparatus}.  While there is a notable degree of initial outgassing from the Viton pieces and the epoxy used to construct the vibration stack and the cavity, the vacuum ion gauge nearest to the cavity reports a pressure of $1\times10^{-10}$ Torr and the lifetime of our atomic cloud in the ODT within the cavity is 7~s.  (The pressure in the production chamber is $2.5\times10^{-11}$ Torr, several times lower than in the cavity science chamber.)  This amount of time is sufficient for evaporating the already ultracold gas to quantum degeneracy, as described below, and  for studying multimode cavity QED physics.  Viewports above and below the cavity mirrors provide optical access to the cavity mode.    Locking light, probe light, and trapping light are coupled to the cavity through these viewports.  We also detect the transmission of cavity fields through these viewports with either CCD cameras, single photon counters,  or photodiodes.  The upper viewport is 49~cm from the cavity mirror:  To prevent the rapid divergence of the beam we  installed an in-vacuum coupling lens to approximately collimate the vertical cavity output.  The lower viewport is sufficiently close to the bottom mirror to render an in-vacuum lens unnecessary.

Alignment of the mirrors for optimal multimode operation is performed before  insertion of the cavity mount into the UHV chamber.  The inset in figure~\ref{cavitychamber} shows the cavity outside the UHV chamber.   We first epoxy the lower mirror to the 5-mm-long cylindrical single-crystal fast piezo, which itself has been  epoxied to the SS mount connecting to the Macor base of the cavity mount.  We next align the upper mirror to this fixed lower mirror.
To optimize the angle and transverse position of the upper mirror, the edge of its flat side is first epoxied to the end of a 3.9 cm-long SS cylinder possessing an inner diameter sufficiently large so as not to obscure cavity transmission.  This SS cylinder is then slipped through a 1.5-mm-larger-diameter mounting ring.  This ring is screwed onto the face of the vertically mounted slip-stick piezo actuator.  The opposite end of the SS cylinder is temporarily held by a positioning stage, and the gap between it and the mounting ring allows us to adjust both the 3-axis position and angle of the mirror while the epoxy filling the gap between the cylinder and ring sets.  For maintaining proper alignment, we monitor cavity transmission resonances at 780~nm while the epoxy sets, adjusting the positioning stage as needed.   We ensure optimal alignment of the mirror with respect to the lower mirror with this procedure, maximizing mode density at the critical length for multimode operation while ensuring good coupling as the slip-stick actuator tunes the cavity toward and away from the $L=R$ condition.

In-vacuum length stabilization of the cavity is achieved by  deriving an error signal using the Pound-Drever-Hall method~\cite{drever_laser_1983,BlackPDH} with 1560-nm light resonant with the TEM$_{0,0}$ mode of the cavity.  Fast feedback is applied to the small single-crystal piezo on the lower mirror. Slow, but large-dynamic-range feedback is applied to the piezo stack in the slip-stick translation stage that actuates the upper mirror. Measurements of the vibration characteristics of the cavity were obtained by loosely locking an optical interferometer to the flat backfaces of each mirror. The amplitude and phase excursions of the interference fringes, when driven by a swept sine-wave, provided an estimate of the transfer functions of the two actuators. Two prominent features were found: a 120-Hz mechanical resonance originating from the slip-stick translation stage and a 3.5-kHz dispersive resonance arising from the coupling of the single-crystal PZT and its mount. An electronic feedback controller was designed based on these observations, incorporating an integrator at DC and open-loop DC gain only limited by the constituent electronic components. Two additional poles in the transfer function between 10~Hz and 100~Hz serve to lower the gain profile rapidly over the next two decades. Two zeros in the transfer function near 1 kHz recover phase margin and result in a stable unity gain point near 10~kHz. The electronic controller also incorporates active resonant filters to provide additional gain to combat the noise induced by the resonance at 120~Hz and to provide the appropriate phase and gain to cancel the 3.5~kHz dispersive resonance. The combination of these elements allows us to achieve  length stabilization with RMS deviation of the cavity length of 1.8(2)~pm [71(5)~kHz at 780 nm] as measured in via the in-loop error signal. This results in a cavity stability nearly 4$\times$ smaller than the intrinsic cavity linewidth $2\kappa$ at 780 nm, whose measurement we provide in the next section.

Light at the wings of the 1064-nm 9.8-W ODT transport  beam ``tweezer" heats the cavity when this beam is focused 40-cm away in the ultracold atom production chamber.  The heating ceases after the focus is translated closer to the cavity, but in the meantime, thermal expansion causes the cavity length to expand.  The length can expand further than the dynamic range  of the fast piezo, at which point cavity lock is lost.  To counter this drift, we complement the fast piezo feedback  with  an extremely slow feedback---unity gain point of 1 Hz---to the piezo stack internal to the slip-stick positioning stage.  This prevents  the DC  component of the fast piezo feedback voltage from railing.  These two feedback loops provide robust locking of the cavity length throughout the 16-s BEC experimental cycle over the course of an afternoon.

To the best of our knowledge, only one other in situ adjustable-length high-finesse cavity has been built for cavity QED research, though for single mode operation~\cite{Rempe_studentthesis_Koch11}.  Strongly coupled cavity QED with a family of near-degenerate higher-order modes and a single atom was explored in reference~\cite{horak}, while a confocal cavity with thermal atoms was explored in references~\cite{Vuletic03_near,Vuletic03_near,vuletic:prl}.

\subsection{Cavity finesse, mode structure,  astigmatism, and birefringence}

The cavity finesse was measured by means of ring-down spectroscopy of the cavity transmission~\cite{Hood01}.  The decay times obtained from these measurements, in conjunction with the cavity lengths obtained from measurements of $\Delta_\text{FSR}$, reveal a finesse of the TEM$_{0,0}$ mode at 780 nm to be 5.72(2)$\times 10^4$.  Similar measurements reveal the finesse at 1560 nm to be 1.22(1)$\times 10^4$ for the same TEM$_{0,0}$ resonance.  The resulting $\kappa$'s are $2\pi\times$~132(1)~kHz and $2\pi\times617(3)$~kHz, respectively. 

We probe the cavity by locking it to a weak 1560-nm laser coupled to a  TEM$_{0,0}$ resonance while addressing the cavity with 780-nm light produced by doubling this same laser. Using the slip-stick translation stage, we measured transmission spectra in the three regimes of intermode spacing of interest to our future cavity QED experiments: confocal degeneracy of one family of modes, mode separation larger than $2\kappa$ and on the order of an atomic linewidth, and modes far apart from one another and satisfying the single-mode criterion $| \Delta_{C} |\ll \Delta\omega_{\{l+m=0\}-\{l+m=1\}} $
 with $\Delta_{C} \leq 10$ MHz.  The results of these measurements are shown in figure~\ref{modetrain}, and the higher-order mode peak heights vary  due to imperfect mode-matching of the input mode, which nominally has a Gaussian waist slightly bigger than the spot size of the TEM$_\text{0,0}$ mode at the mirror face.  The use of the locking laser as the source for the probe laser ensures that the cavity remains stable with respect to the probe light despite the potential drift of the locking laser.  The frequency drift of this narrow, temperature-stabilized fiber laser is slow and fractionally insignificant in the large atomic detuning limit.

Under the paraxial approximation, the transverse mode frequencies of a Fabry-P\'{e}rot cavity made from parabolic mirrors are
\begin{equation}\label{eqn1}
f_{nlm}=\frac{c}{2L}\left(n+\frac{m+1/2}{\pi}\theta_m    +\frac{l+1/2}{\pi}\theta_l\right),
\end{equation}
where $\theta_i=\arccos{\sqrt{(1-\frac{L}{R_{i1}})(1-\frac{L}{R_{i2}})}}$ and $R_{ij}$ is the radius of curvature of the $j$th mirror in the $i$ direction~\cite{siegman}.  The ideal confocal cavity corresponds to the condition  $\theta_l = \theta_m = \pi /2$ such that the resonance frequency is unchanged as long as $n + (m+l)/2 = \text{constant}$. 

In practice, however, highly polished mirror substrates are spherical rather than parabolic, and this spherical aberration reduces the number of supportable higher-order modes~\cite{Vuletic01}.
Moreover, mirror substrates may not be  cylindrically symmetric due to mounting stress or the opposing mirrors may be mounted off-axis and at a relative angle.  Consequently, they exhibit astigmatism, meaning that the cavity has different radii of curvature along different transverse axes. In this case $\theta_l \neq \theta_m$, with, say, $\theta_m$ corresponding to the direction with  maximum  $R$ and $\theta_l$ corresponding  to the direction with minimum  $R$, or vice-versa.  In such cases, it is no longer possible to ensure that $f_{nlm}$ only depends on $n + (m+l)/2$: the two transverse degrees of freedom are no longer equivalent. The astigmatic cavity then exhibits two confocal degeneracy points: one with degenerate $l$ modes  and $\theta_l = \pi/2$, and one with  degenerate $m$ modes  and $\theta_m = \pi/2$. Equation~\ref{eqn1} may now be rewritten in terms of a frequency shift proportional to $m+l$ combined with a correction along one of the transverse axes:
\begin{equation}
f_{nlm } = \frac{c}{2L}\left( n + \frac{m+l}{\pi}\theta_l   +  \frac{l}{\pi} \xi + \frac{2\theta_l + \xi}{\pi}    \right),
\end{equation}
where $\theta_m = \theta_l + \xi$. 

Our cavity is found to display a small amount of astigmatism---perhaps due to mounting stress or imperfections in the manufacturing---with major and minor axes located at $\sim$45$^\circ$ to the science chamber principle axes; see figure~\ref{cavitychamber}. The two radii of curvature, measured by the length of the cavity at the two degeneracy conditions, are $R_1 = 9.959(1)$~mm and $R_2 =  R_1 + 8.6(1.1)$ $\mu$m.  The respective spectra for these lengths are shown in figure~\ref{modetrain}(b) and (c). They are characterized by a single peak of nearly degenerate modes next to a broad shoulder of modes extending out to either lower or higher frequencies. The modes around the sharp peak are associated with the family of modes satisfying the degeneracy condition of one of the astigmatic axes, while the broad pedestal of nearly degenerate modes to the side is associated with the  unmet degeneracy condition of the other astigmatic axis. Changing the cavity length from that shown in figure~\ref{modetrain}(b) to figure~\ref{modetrain}(c) reverses the family of modes---$l$ versus $m$---that satisfies the degeneracy condition.  The dispersion of each family of modes is likely due to mirror misalignment~\cite{ChanThesis03}.

The cavity is effectively single mode at a large distance from confocallity, and the difference between the major and minor radii of curvature becomes negligible compared to $L-R$.  The astigmatism then appears as a dispersion of spectrally isolated families of modes with constant $l+m$, as in figure~\ref{modetrain} (e). We note that much larger splittings between families of modes are possible via further translation of the positioning stage.

The maximum number of degenerate modes in the cavity is limited by the mirror imperfections which cause the finesse to decrease with increasing transverse mode index and by spherical aberrations which introduce non-linearities in $f_{nlm}$.   We will assess in future work where the limit in the number of supportable modes lies for this cavity.  However, we can obtain a rough estimate by noting the largest index mode we image on a CCD camera at a length just shy of confocallity.  Unlike for a concentric cavity, the waists of the modes of a confocal cavity do not diverge at the mirror upon reaching the $L=R$ degeneracy point, and so we can assume that any mode supported at $L=R\pm\epsilon$ will likely remain so at $L=R$.  We find that $l$ and $m$ of up to order 50 each may be observed, providing a rough estimate of $lm/4 \approx 10^3$ as the number of same-parity modes supported within a $\sim$50-MHz bandwidth in figures~\ref{modetrain} (b) and (c). 

Birefringence is another aspect of non-ideal cavity behavior. The cavity exhibits no measurable birefringent splitting between orthogonally polarized light,  and is therefore capable of supporting both linearly and circularly polarized modes.

\subsection{Cavity dispersion}\label{dispersionSec}

The high reflectivity of the cavity mirrors is obtained via a multilayer dielectric stack which acts as an interference filter. The characteristics of this filter, unlike a simple metal mirror, are strongly wavelength dependent~\cite{Hood01}. The penetration depth of light into the dielectric stack is different at 780~nm versus 1560~nm.  As a result, the effective length of the cavity is different at these two wavelengths. To first order in small changes $\delta L$ of the cavity length, and ignoring the Gouy phase, the line shift of the cavity obeys the following relation: $\delta\nu / \nu = \delta L/L$ with $\delta\nu_{780} = 2 \times \delta\nu_{2\times780 = 1560}$.  This is true in the absence of any dispersion in the effective cavity length.  However, real  cavity coatings exhibit dispersion and $L_{1560} = L_{780} + \Delta L$, where $\Delta L$ is a required correction to obtain the frequency relationship between  cavity resonances at the two wavelengths: $\delta\nu_{780} = (2+ \epsilon_D) \delta\nu_{1560}$. In addition to modifying the behavior of the cavity with respect to length changes, the dispersion of the cavity length also results in an absolute frequency offset $\Delta_D$ between the frequencies at which the 1560-nm laser and the doubled light are resonant with the cavity. In order to have resonant probe light at 780~nm and locking light at 1560~nm, sidebands must be introduced onto the 1560~nm laser at either $\Delta_D$ or $\Delta_D - \Delta_{FSR}/2$.

We find the factor $\epsilon_D$ by varying the frequency of the locking sideband required to address the cavity resonance at $1560$-nm and then measuring the resultant translation of the $780$-nm cavity resonance when the cavity is locked at this new frequency: $\epsilon_D = 5.83(5)\times 10^{-4}$. This information, together with the laser frequency and $\Delta_D$, provides sufficient information to infer that the effective length difference $\Delta L= 2.78(3) \ \rm{\mu m}$. This length difference is beneficial for locking because the cavity is not simultaneously degenerate for both wavelengths. Instead, the 1560-nm resonances are separated by $\sim$4.5$\kappa$ when the cavity is degenerate at 780~nm, at the $L = R_2$ condition.

\section{Ultracold gas production, transport, condensation and intracavity trapping}\label{BEC}

The creation of ultracold gases in the production chamber follows the work of Lin \textit{et al.}~\cite{Lin09}. 
An atomic beam, decelerated by a Zeeman slower~\cite{MetcalfBook99}, loads $^{87}$Rb into a magneto-optical trap (MOT) at the center of the production chamber's trapping region.  See figure~\ref{apparatus}.  
Subsequently, the atoms are  optically pumped to the $\ket{1,-1}$ state.  
Quickly thereafter, we increase the MOT's magnetic quadrupole field to obtain a gradient of 195 G/cm.  This traps approximately one-billion atoms in the $\ket{1,-1}$ state. Linearly ramping the RF frequency in coils external to the chamber from 25.5~MHz to 5.5~MHz over 5.7~s produces an evaporatively cooled cloud of $8\times10^7$ atoms at 40~$\mu$K.
An off-resonant ODT is tightly focused below the center of the quadrupole trap, and the atoms are directly loaded into this ODT during the course of the evaporation.  We use this ODT as an optical tweezer for transporting the atoms into the cavity science chamber. It employs 9.8~W of 1064-nm laser light focused to waists of 50~$\mu$m and 73~$\mu$m in the horizontal and vertical directions, respectively. Loading into the ODT from the magnetic trap is performed by significantly reducing the quadrupole gradient over 1~s to a value just supporting the atoms against gravity. In this configuration, the quadrupole trap provides confinement in the weakly trapped, axial direction of the ODT beam.  By lowering the optical power, the atoms can undergo efficient optical evaporation to high phase-space density.  We have obtained a BEC of $1\times10^6$ atoms at 200~nK in the production chamber by further reducing the ODT power. 

\begin{figure}[t!]
\centering
\includegraphics[width=0.5 \columnwidth]{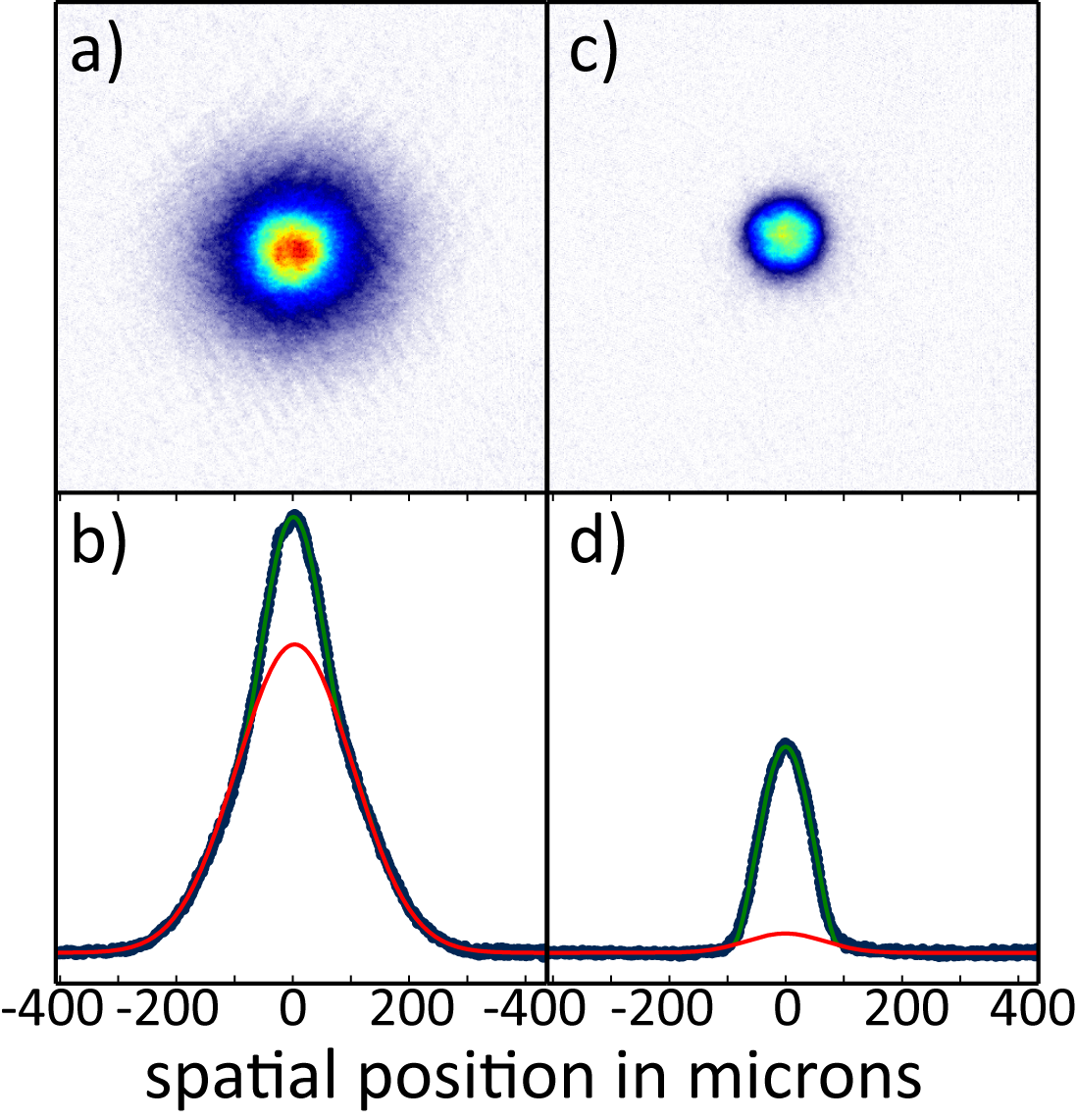}
\caption{Typical single-shot absorption imaging density profiles of $^{87}$Rb atomic clouds after 18~ms of free expansion. The top row shows the density distribution, with the colorscale equal for the two pictures. The lower row shows the integrated density along a direction perpendicular to the cavity axis. The black dots represent the data, the green curve shows the fitted bimodal profile and the red curve shows the fitted thermal portion. (a) and (b): $1.8(2)\times10^5$ atoms at the onset of BEC, $T\sim320$ nK.  (c) and (d):  Nearly pure BEC of $4.5(5)\times10^4$  atoms at $T\approx150$ nK. BEC fraction is $\sim$90\%.} \label{BECpics}
\end{figure}
However, creating BECs before ODT transport to the cavity science chamber is counter-productive, as three-body collisions heat the gas during transport~\cite{Ketterle01}.  Rather, we evaporate to 8~$\mu$K above T$_\text{c}=1.6$~$\mu$K and transport $6\times10^6$ atoms to the cavity science chamber by translating the last focusing lens of the ODT.  This lens is mounted on an air bearing translation stage.  The stage provides a translation of the lens and  of the ODT focus over a distance of 35~cm in 4.3~s.  The position of this beam is adjusted to maximize loading of atoms into the ODT while maintaining a transport trajectory whose final location is the center of the multimode cavity.  The motion of the stage is chosen to be a smooth function defined by three sections of constant acceleration, with the central section having zero acceleration. The parameters of this curve were chosen so as to maximize transfer of atoms into the science chamber. The overall transfer efficiency is 30\%.

The two sections of the apparatus, the ultracold atom production chamber and the cavity science chamber, are separated by a gate valve, allowing us to break vacuum in one without disturbing the other.  This feature will prove useful for modifing the cavity, if for instance we want satisfy the concentric criterion $L=2R$ or exchange the mirrors with ones of differing finesse or $R$.

\subsection{BEC production within the cavity}

Once the atomic cloud is positioned between the mirrors in the science chamber, a second 1064-nm laser beam crosses it to confine the atoms. This beam is circular with a Gaussian waist (1/e radius) of 48~$\mu$m.  Forced evaporation  in this crossed ODT produces a nearly pure BEC of $4.5(4)\times 10^5$ atoms at density $4.7(5)\times10^{14}$ cm$^3$, as shown in figure~\ref{BECpics}.  The trap  frequencies are 145(5) Hz, 115(1) Hz, and 167(2) Hz in $\hat{x}$, $\hat{y}$, and $\hat{z}$, respectively. We expect to achieve pure condensates with 10$^5$ atoms after performing additional evaporation optimization.  The total time to achieve a BEC within the cavity, starting from MOT loading, is 16 s.

\subsection{Intracavity FORT}

For ease of coupling to the cavity mode at 780-nm, we can also load a thermal gas into an intracavity FORT formed with the 1560-nm locking light in the TEM$_\text{0,0}$ mode. To do so, we increase the power of the locking laser until the steady state circulating power provides a trap depth of 5.2(1.3)~$\mu$K. The transport ODT brings atoms into the science chamber and holds the cloud within the cavity mode. Addition of a second  ODT beam crossing the first increases the mode matching between the 1064-nm trap and the 1560-nm cavity mode. Optical evaporation with the crossed 1064-nm beams loads atoms in the intracavity FORT, resulting in $2.5(2)\times10^5$ atoms at 1.8(5)~$\mu$K confined in the 1560-nm standing-wave TEM$_\text{0,0}$ cavity mode. The temperature and atom number uncertainties, as well as the relatively high final temperature, are dominated by the lack of intensity stabilization and dynamical intensity adjustment of the FORT power, all of which will be added in the near future.  

\begin{figure}[t!]
\centering
\includegraphics[width=0.8 \columnwidth]{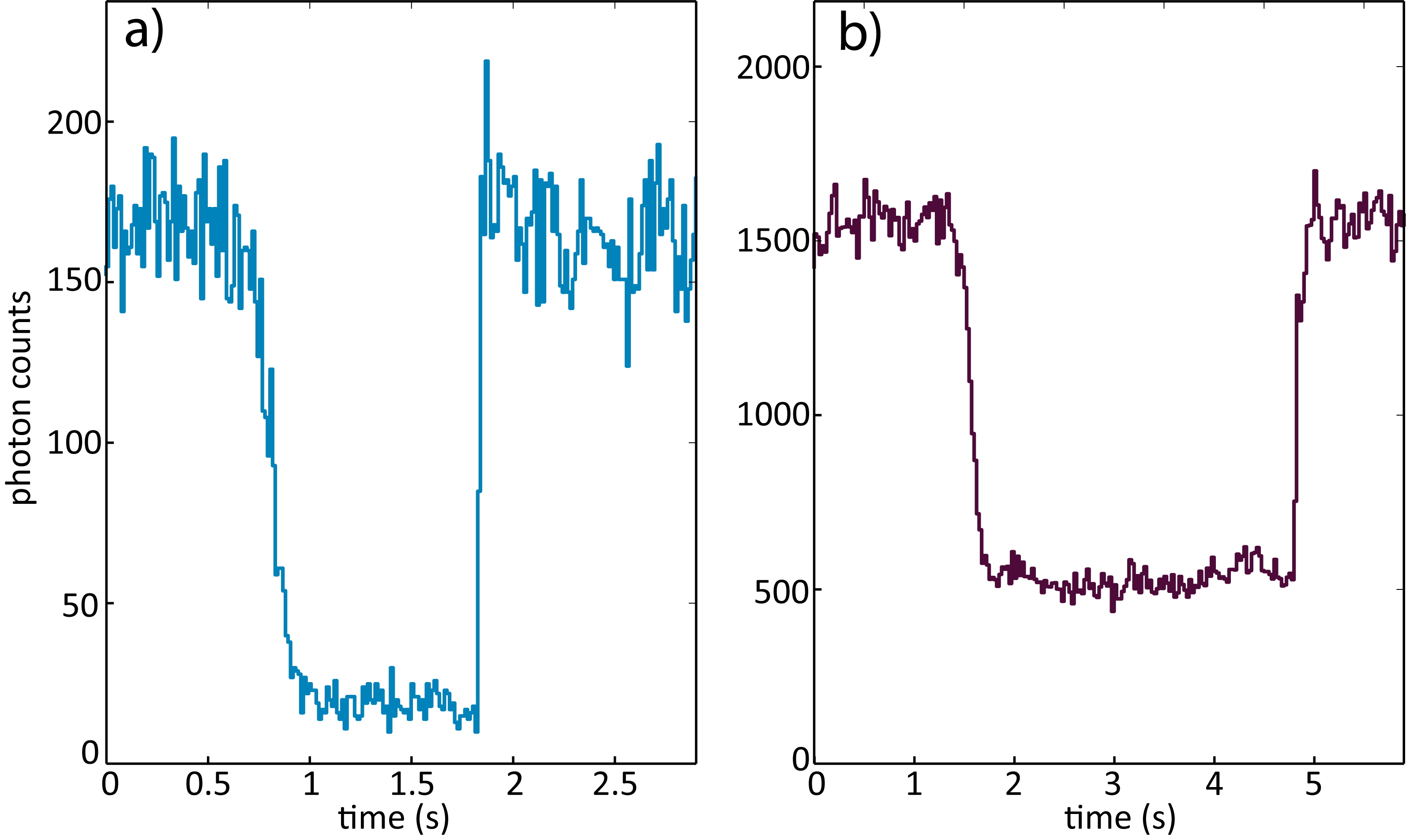}
\caption{Cavity transmission versus time as atoms are transported into the cavity by the ODT tweezers beam.  a) Single-mode regime  as shown in figure~\ref{modetrain}e and $\Delta_A \equiv \omega_c - \omega_a  = -7.5(1) \ \rm{GHz}$ with bin width 12.5~ms, and b) multimode regime as shown in figure~\ref{modetrain}c with $\Delta_a = 31.3(1) \ \rm{GHz}$ and bin width 25~ms.} \label{transits}
\end{figure}

The atoms in the FORT are thermal and the in-trap density distribution obeys Boltzmann statistics with $\rho(\mathbf{x}) = \exp(-V(\mathbf{x})/T)$, where $V(\mathbf{x})$ is the standing-wave optical potential of the FORT. The resulting atomic density distribution is Gaussian, assuming a harmonic approximation to the potential.  Such an approximatin is valid as long as the atoms are sufficiently cold so as to localize close to the potential minima. We choose to operate at larger trap depth, though lower atom number, in the experiments discussed below.  Since the atoms are loaded into the FORT from an ODT with waists larger than $\lambda/2$, the atoms will occupy multiple wells of the FORT.   The waist of the 1560-nm TEM$_{0,0}$ cavity mode is $\sqrt{2}$ larger than the 780-nm TEM$_{0,0}$ mode waist: the FORT waist is $35\sqrt{2} = 50$~$\mu$m.  For the trap parameters used below---$V_{FORT} =22(6)$~$\mu$~K and $T = 5(1)$~$\mu$~K---the atomic density distribution has a transverse waist of $\sim$17~$\mu$m, which is smaller than the 35~$\mu$m waist of the 780-nm mode.

\section{Atomic ensemble-cavity coupling in single-mode and multimode configurations}\label{coupling}

We now present results demonstrating the coupling of an atomic ensemble trapped in the intracavity FORT to the cavity, both in the multimode and in the single-mode configurations.

\subsection{Atomic ensemble transiting cavity mode(s)}

The first measurement is of the transmission of the cavity as atoms are transported into the cavity by the ODT tweezer.  As shown in figure~\ref{transits} for both the single mode and the mulitmode configurations, at $t=0$, light is injected into the cavity on resonance and the transmission is recorded on a single photon counter.  As the atoms enter the cavity, the atomic ensemble-cavity interaction dispersively shifts the cavity line out of resonance with the probe laser and the transmission drops~\cite{Kimble1998,Hood00,CohenTannoudji}. After a hold time, the atoms are removed from the cavity by resonant imaging transverse to the cavity axis, and the transmission returns to the original value.  The detection efficiency by the single-photon detector is 10\% for the single-mode configuration and in the multimode case, $\sim$10\% for the lowest order modes.  This efficiency decreases for  higher order modes due to poor mode-matching into the detector, but will be improved in the near future.

\begin{figure}[t!]
\centering
\includegraphics[width=0.8 \columnwidth]{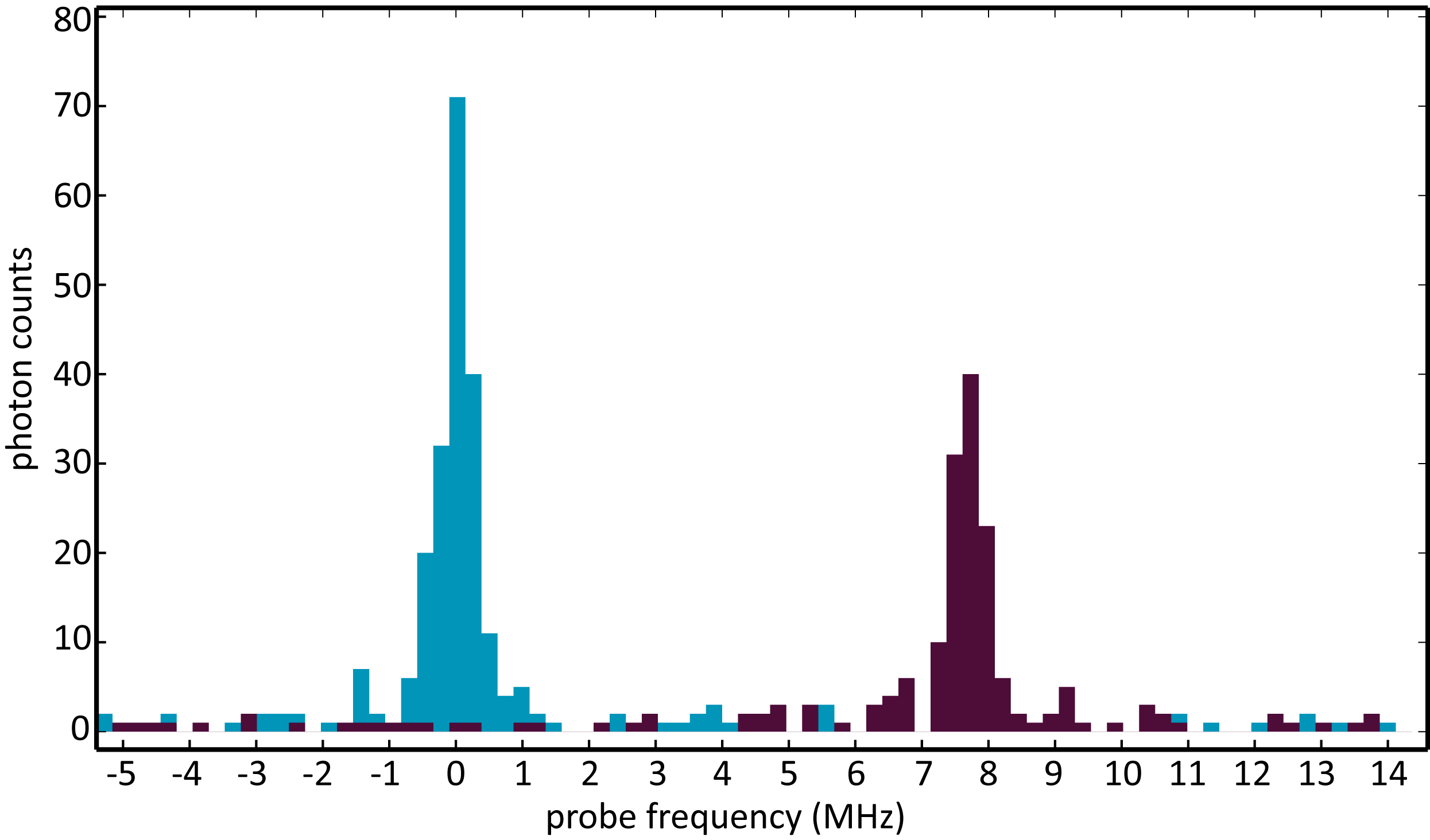}
\caption{Single-shot data of bare cavity resonance (left, blue) and dispersively coupled cavity resonance (right, purple) versus frequency for a single-mode configuration with $L >R$, as shown in figure~\ref{modetrain}e.  The coupled resonance is shifted by $\delta\omega^{SM}_c = 7.66(2)$~MHz. Data taken for $N = 2.0(2)\times10^5$ atoms and $\Delta_A = \omega_c - \omega_a = 8.7(1) \ \rm{GHz}$.
 The sweep rate is 20~MHz/40~ms.} \label{SM}
\end{figure}
\subsection{Single-mode cavity dispersive shift}

In a second series of experiments, we measure the dispersive shift of the cavity resonance with atoms confined within the intracavity FORT and with a  pump field injecting 1.7(3) photons when on resonance with the cavity.  In the single-mode cavity case of figure~\ref{SM}, we observe that coupling of the atomic ensemble to the cavity  shifts the cavity resonance by $\delta\omega_c = 7.66(2)\ \rm{MHz} \gg \kappa$, as expected from considering the normal-mode spectrum of the Tavis-Cummings model of single-mode, multi-atom cavity QED~\cite{Kimble1998,CohenTannoudji}. Using numerical simulations of the optical Bloch equations~\cite{CohenTannoudji}, we determine that this polariton shift is only $\sim$1\% less than what it would be if a much weaker probe were used.

We may use the single-mode cavity dispersive shift to obtain a measurement of the single atom coupling constant $g_0$ via  $\delta\omega^{SM}_c = N_\text{eff} g_0^2/\Delta_A$, where $N_\text{eff} = \alpha_\text{0,0} N$ is the effective number of atoms coupled to the cavity TEM$_\text{0,0}$ mode and $\Delta_A = \omega_c - \omega_a$.  The overlap $\alpha_\text{0,0}$ of the atomic cloud confined within the intracavity 1560-nm standing-wave FORT with the electric field of the 780-nm cavity standing-wave is given by~\cite{Barrett12}: 
\begin{equation}\label{overlap}
\alpha_\text{0,0} = \frac{1}{2}\frac{1+e^{-4/\eta}}{1+2/\eta} = 0.5(1),
\end{equation} with $\eta = V_\text{FORT}/k_B T$.  The trap depth $V_\text{FORT}$ of our FORT is 22(6)~$\mu$K and the atoms are at $T=5(1)$ $\mu$K.  We determine that $g_0=2\pi\times1.0(1)$~MHz on the $\ket{5S_{1/2},F=2,m_F=-2}\rightarrow\ket{5P_{3/2},F=3,m_F=-3}$ cycling transition or $g_0=2\pi\times7.2(8)\times 10^5$~Hz on the $\ket{5S_{1/2},F=1,m_F=-1}\rightarrow\ket{5P_{3/2},F=2,m_F=-2}$ transition.  The expected values of $g'_0 = 2\pi\times 1.47$~MHz and $g'_0 = 2\pi\times1.04$~MHz respectively based on the measurements of the cavity $R$ and $L$ mentioned above. The analytical expression  for the coupling is $g'_0 = \sqrt{\mu^2\omega_c /\hbar\epsilon_0 V}$, where $\mu$ is the transition matrix element, $V=\pi w_0^2 L/4$ is the mode volume and $w_0=\sqrt{R\lambda/2\pi}=35$ $\mu$m is the mode waist ($1/e$ mode radius). These estimates are based on the measurements $N = 2.0(2)\times10^5$  and $\Delta_A = 8.7(1) \ \rm{GHz}$, in addition to the use of the appropriate transition matrix elements for the D2-line of  $^{87}$Rb at this detuning, the polarization of the pump beam, and the direction of the quantization axis set by an applied magnetic field.  We note that the single-atom cooperativity when driving on the $\ket{5S_{1/2},F=2,m_F=-2}\rightarrow\ket{5P_{3/2},F=3,m_F=-3}$ transition is $C\approx 2.5$.

\begin{figure}[t!]
\centering
\includegraphics[width=0.8 \columnwidth]{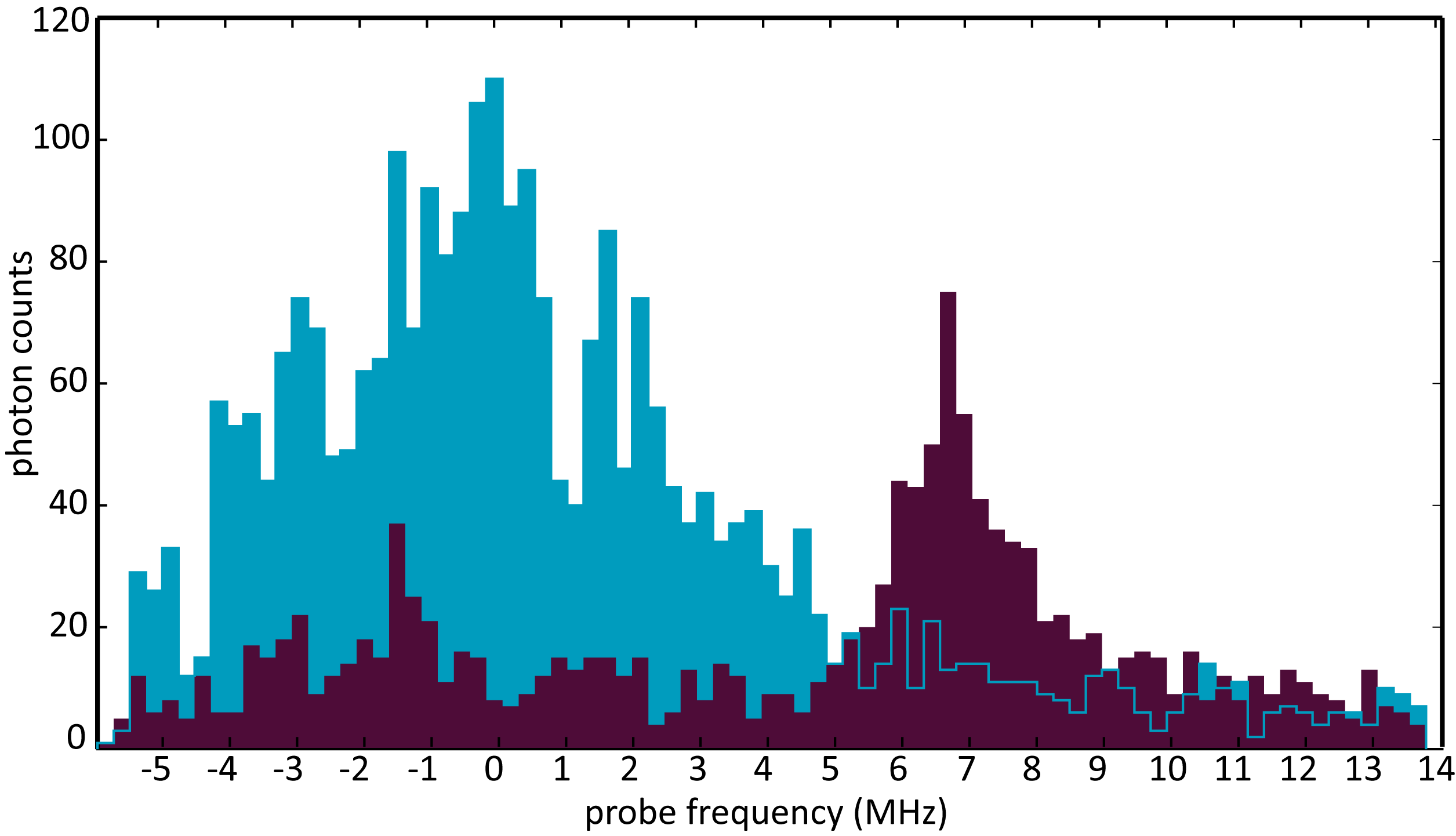}
\caption{Single-shot data of bare cavity resonances (left, blue) and dispersively coupled cavity resonances (right, purple) versus frequency for the multimode configuration with $L=R_2$ (see figure~\ref{modetrain}c), taken for $N = 2.2(2)\times10^5$ atoms and $\Delta_A = \omega_c - \omega_a = 31.3(1) \ \rm{GHz}$. By fitting a broad Gaussian to each, we find that the coupled resonances are shifted by $\delta\omega^{MM}_c = 7.1(2)$ MHz.  Note that $\Delta_A$ differs from the single-mode cavity data in figure~\ref{SM}.    The sweep rate is 30~MHz/250~ms.} \label{MM}
\end{figure}

The disparity between the measured $g_0$ and the corresponding $g'_0$ is likely due to an overestimate of $\alpha_\text{0,0}$.  The harmonic approximation to $V(x)$ is not quite accurate for these small ratios of trap depth-to-temperature.  A breakdown of this approximations would tend to lower  $\alpha_\text{0,0}$, which in turn would increase the $g_0$ estimate closer to $g'_0$. A more accurate calculation of $\alpha_\text{0,0}$, together with improved  stability of the FORT power and better measurements of the intratrap $T$ and $N$ of the atoms, will allow us to obtain a better value for $g_0$.

\subsection{Multimode cavity dispersive shift}

The atomic ensemble-cavity coupling induces a change in both the shape and center frequency the dispersively shifted portion of the cavity transmission spectrum in the multimode case, as shown in figure~\ref{MM}.  One may use such data to measure an effective coupling $g_\text{eff}$ to the cavity modes, and use this parameter to then estimate the number of modes coupled to the atomic ensemble~\cite{Vuletic01,Ritsch02_trans,lev}.  However, such an analysis  for a longitudinally (directly) pumped cavity QED system is complicated by two sets of spatial overlaps: One must consider the overlaps of both the pump spatial mode and the atomic ensemble distribution with the cavity modes. Only the latter needs to be considered in transversely pumped cavities, i.e., those pumped via the Rayleigh scattering of a plane-wave off the atoms. We are in the process of carrying out such an analysis---including an analysis of the spatial structure of out-coupled light~\cite{ChanThesis03}---for both the longitudinal and transversely pumped configurations.  We will report in a future publication resulting estimates for the number of modes coupled to the atomic ensemble.

\section{Prospects for beyond mean-field physics with multimode cavity QED}\label{prospects}

We now describe some of the  exotic many-body states that may be accessible with BEC-multimode cavity QED systems, as proposed in references~\cite{Gopalakrishnan09,Gopalakrishnan10,Gopalakrishnan:2011cf,Gopalakrishnan:2011jx,Strack:2011hv,Muller12,Buchhold13}.   These systems consist of a cavity pumped indirectly by  coherently scattering light off intracavity atoms~\cite{Vuletic01,Vuletic03_near,vuletic:prl,ritschprl,ritschpra}. 
As mentioned in section~\ref{intro}, beyond mean-field physics may play a role in the system organization due to the fact that multiple degenerate modes of the cavity support localized fluctuations.   To qualitatively explain this in more detail, we first describe self-organization of atoms within a single-mode cavity before following the analyses given in references~\cite{Gopalakrishnan09,Gopalakrishnan10}.   

The infinite-range atom-atom interaction mediated by the off-resonant scattering of pump photons induces an instability in the otherwise homogeneous density of the gas above a critical threshold intensity of this pump field~\cite{vuletic:prl,ritschprl,ritschpra}.    The atoms are liable to organization with $\lambda$-periodicity in one of the two checkerboard patterns formed by the interference of the pump and cavity fields.  
While the amplitude of both the cavity field and the modulated atomic density emerge due to the light-matter interaction, the spatial structure is predetermined by the cavity geometry.  The atoms have only one of two ways to organize and no small deformation of the atomic arrangement connects the two patterns.  The process of self-organization is therefore one of discrete symmetry-breaking.   While this dynamical instability to self-organize may be considered a (second-order) phase transition, the organized phase does not possess the rigidity or compliance of a true crystalline solid:  there are no phonon-like excitations of small deformations.  This second-order mean-field transition in a single-mode cavity is heralded by superradiance and may be described in the framework of the Tavis-Cummings model or the Hepp-Lieb-Dicke model~\cite{vuletic:prl,ritschprl,ritschpra,Charmichael07,morigi08,domokos08,Nagy10,morigi10,Keeling10,baumann10,Baumann11,Mottl12,Barrett12PRL,Brennecke13,Barrett14PRL,Hemmerich14}. Superradiance arising from atomic density instabilities has also been observed in transversely pumped ring cavities~\cite{slama,Zimmermann14} and for thermal gases in free space~\cite{Gauthier12EPL,Gauthier12PRA}.  A thermal gas pumped near a mirror has also been shown to exhibit optomechanical self-structuring~\cite{Ackemann12,Labeyrie:2014gh}.

The situation of a transversely-pumped multimode cavity is quite different.  The atomic density may self-organize into the checkerboard pattern of any mixture or coherent superposition of the large number of degenerate modes.  So in addition to the discrete $\mathbb{Z}_{2}$ symmetry of the $\lambda$-period lattice, a quasi-continuous symmetry is broken.  This symmetry may best be viewed in a quasimomentum picture~\cite{Gopalakrishnan09,Gopalakrishnan10}:  The density of the atoms $\rho$ may be decomposed into the basis of the TEM$_{l,m}$ mode functions; the mode indices $l,m,n$ now serve as the axes of a quasimomentum space, where $n$ is the number of longitudinal nodes along the cavity axis. For a gas confined in a 2D plane spanned by the cavity axis and the pump field wavevector, we may write $\rho = \sum c_{m,n} \rho_{n,m}$, where we have taken the direction associated with $l$ to point perpendicularly to this plane.  It may be shown that the ``soft'' modes  the atoms prefer to self-organize within satisfy an equation of quasimomentum conservation $n+(l+m)/2=k_{0}R$ for a confocal cavity $L=R$~\cite{Gopalakrishnan09,Gopalakrishnan10}. ($n+l+m=k_{0}R$ for a concentric cavity $L=2R$.) This equation is none other than the characteristic equation for the axial-plus-transverse modes of a Fabry-Per\'{o}t cavity, equation~\ref{eqn1}~\cite{siegman}. 

For $l=0$, the soft modes span two lines in the $n-m$ quasimomentum plane, intersecting at $n =k_{0}R$.  The $\mathbb{Z}_{2}$ symmetry is represented in the reflection of these lines about the $m$-axis, i.e., in the $-n$ portion of the plane.  (The validity of allowing  this index to be negative is more clear when using a Laguerre-Gaussian mode basis.)  Dominant interactions exchange quasimomentum among these sets of points.  Energetically costly interaction fluctuations exchange quasimomenta located above this 2D-plane via  modes with non-zero $l$.  These fluctuations are sufficiently strong so as to renormalize the Landau coefficients in the system's free energy, converting what was a second-order phase transition into a weakly first-order transition, as determined from a functional integration analysis  and following the renormalization group program laid out by Brazovskii for an analogous, classical soft matter system~\cite{Brazovskii:1975wq,Hohenberg:1995gm}.  

This intriguing quantum Brazovskii transition, a rare example of a fluctuation-induced first-order transition, is very much a non-mean-field transition.  Not only is the threshold of the transition renormalized upward, but the actual character of the transition---first versus second-order---differs from the single-mode case.  Moreover, this weakly first-order phase transition is driven by both quantum and classical fluctuations, unlike the single-mode case, which only is  liable  to classical noise. (Quantum fluctuations diminish as the atom number $N$ increases in a single mode cavity~\cite{Strack:2011hv}.) Indeed, while interactions in the single-mode case are infinite-ranged, the fluctuations in the quasimomentum space are limited to 0D points given by $\pm n = k_{0}R$.  The space liable to these fluctuations in the multimode case is much larger, occupying the 1D lines delimited above.   As the number of degenerate modes increases, the lines the soft modes trace form a quasi-continuous distribution in quasimomentum space:  Breaking this quasi-continuous symmetry induces Goldstone modes in the ordered phase that manifest as phonon-like excitations of the atomic positions.  The order exhibited by this quantum Brazovskii transition is quite unusual---a superfluid smectic quantum liquid crystal---and worthy of exploration both in its own right and due to the relevance of these to technologically relevant strongly correlated materials~\cite{Gopalakrishnan09,Gopalakrishnan10,Simons10,Fradkin12}.  Geometrical frustration destroys this quantum liquid crystal   if a stack of 2D BECs are trapped in a lamellar pattern spanning either side of the cavity axis.  The ensuing phase should resemble a superfluid glass~\cite{Boninsegni:2006cp,Gopalakrishnan09,Gopalakrishnan10}.
 
 Each of these manifestations of quantum soft matter may be heralded by telltale time or spatial correlations in the superradiant emission from the cavity mode  as threshold is reached~\cite{Gopalakrishnan09,Gopalakrishnan10}.   Bragg peaks in the atomic momentum distribution emerge in the single-mode case~\cite{Baumann11,Hemmerich14} and are a direct indication that a fraction of the matter wave has self-organized into a periodic pattern.   In the supersmectic phase, these Bragg peaks should broaden into arcs, just as in the case of x-ray scattering off classical  liquid crystals~\cite{chaikin1995principles}.
 
 Fixing the positions of spinful atoms at random locations within the cavity modes using an external, static optical lattice sets the stage for exploring spin-glass physics and the closely associated architectures for neuromorphic photonic computation~\cite{Gopalakrishnan:2011cf,Gopalakrishnan:2011jx,Strack:2011hv,Muller12,Buchhold13,Marandi:2014ik}.  Atoms with one spin state  coupled to a transverse pump field, but with the other state coupled to the cavity mode(s), exchange spin-flip excitations when the Raman-resonance condition is off by a detuning $\Delta_{C}\gg\kappa$.  In a single-mode cavity, this disordered XY-spin model realizes a ``Mattis model'' resulting in a spin-helix texture~\cite{Weissman:1992wr}.  Adding  additional modes to the cavity frustrates the interactions.  With a sufficient number of degenerate modes, the system should no longer support long-range spatial spin order, though order may develop in time, resulting in the aging and arrested dynamics characteristic of a spin glass.  Observations over many decades in time is possible from  the quantum-fluctuation driven to the thermally driven timescales.

\section{Conclusion}

Future work will strive to demonstrate the use of this novel adjustable-length cavity and Bose-Einstein condensate  multimode cavity QED apparatus in searches for these exotic states of quantum matter.  The strong and spatially dynamic light-matter interaction provided by cavity QED  may also be exploited for applications outside the exploration of many-body physics such as quantum information processing~\cite{Kimble:2008uv}, cavity optomechanics~\cite{StamperKurn12} and sensing~\cite{Schreppler27062014}.

\ack{
We thank James Thompson for helpful discussions, Yijun Tang and Shenglan Qiao for early experimental assistance, and Sarang Gopalakrishnan and Paul Goldbart for enlightening theory discussions. We acknowledge support from the David and Lucille Packard Foundation, the DARPA YFA program, the ONR YIP program, and the NDSEG fellowship program.}

\section{References}
\providecommand{\newblock}{}


\end{document}